# On the molecular theory of dimer liquid crystals

P. K. Karahaliou, A. G. Vanakaras* and D. J. Photinos*Department of Materials ScienceUniversity of Patras, Patras 26504, Greece.*

*Author for correspondence, vanakara@upatras.gr, tel, fax: +30 2610 996286

**Abstract**
We present a statistical mechanics approximation scheme for the explicit treatment of spacer-mediated configurational correlations among the mesogenic units that form a dimer molecule. The approximation is applied to the description of the nematic phase of linear uniaxial dimers interacting via a standard molecular pair-potential. Transition temperatures, order parameters and pair correlation averages are calculated for different spacer lengths. The results readily reproduce the experimentally observed trends of phase transition thermodynamics and of dipolar correlations deduced from dielectric studies.

# 1. Introduction

Typical liquid crystal (LC) dimers consist of two mesogenic cores linked by means of a flexible alkyl spacer. Their first appearance in the literature dates back to Vorländer [1]. Dimer LCs reappeared (under various names) in the literature about half a century later [2, 3] and their systematic study was put forward by G.R. Luckhurst and coworkers about twenty years ago [4-6]. Part of the interest in LC dimers is related to their usefulness as model systems for studying LC polymers [3-7] but their polymorphism makes them also interesting in their own [8]. The mesomorphic behaviour of LC dimers shows certain striking differences from that of simple mesogens. In particular, the transition temperatures from the isotropic to the LC phase, and the respective order parameters, of some of the most common homologous series of dimers, show dramatic alternations with spacer length. These alternations are usually referred to as 'odd-even effects', where the parity refers to the number of carbons in the spacer.

Despite the advances in the experimental study of LC dimers [8], molecular theory has not progressed much beyond the initial successful interpretation of the odd-even effects in terms of spacer conformations. More specifically, the influence of the core-spacer linkage geometry on the orientational ordering of the dimers in the nematic phase has been studied by NMR spectroscopy and has been accounted for quite accurately with the cord model [9]. This model was subsequently extended to provide a unified description of orientational ordering and nematic-isotropic phase transitions for linear dimers, oligomers and main-chain polymers [10]. Moreover, various alternative schemes of conveying the structural differences between odd and even dimers have been proposed and explored in the context of phenomenological nematic mean-field calculations [11, 12] and molecular simulation studies [13]. Finally, a systematic description of the shape and deformability of dimers in terms of structural shape parameters has been introduced [14] and applied to the identification of the intrinsic shape parameters that give rise to the odd-even effects in dimers with various core-spacer linkage geometries. Nevertheless, a number of important issues, such as the very rich smectic polymorphism of LC dimers [8], the possibility of their biaxial ordering in the nematic phase, their static and dynamic dielectric properties, remain essentially unaddressed on the molecular theory level. On the other hand, computer simulations [15, 16] have successfully explored smectic polymorphism and microsegregation phenomena in model LC dimers, thus offering new insights that were inaccessible by molecular theory.

The differences between the nematic behaviour of simple mesogens (hereafter to be referred to as monomers) and dimers are thought to stem entirely from the configurational correlations that the spacers impose on the mesogenic units [4-6, 9, 10]. Accordingly, the inclusion of such correlations constitutes the major step in the formulation of a molecular theory that goes from the description of single mesogen self-organization to dimer and then on to multi-mesogen phases. The present work is concerned with this step. To develop a suitable statistical mechanics approximation we formulate the configurational partition function of a system of dimers starting from the mesogen-mesogen pair potential and distinguishing *inter*- from *intra*-dimer pairs, the latter bearing explicitly the constraints dictated by the conformations of the spacer. We then introduce a variational field and apply a pair-wise decoupling approximation to obtain the free energy. The functional minimization of the latter determines the variational field. The general procedure is outlined in section 2 of the paper and is then applied to the formulation of the free energy for the nematic-isotropic phase transition. This statistical mechanics framework is then used in section 3, together with standard forms of the molecular interactions (Gay-Berne potential for the mesogenic cores and Ryckaert-Bellemans torsional potential [16, 17], or alternatively RIS conformation statistics [18], for the spacer), to carry out phase transition calculations and to evaluate the temperature dependence and spacer-length dependence of measurable order parameters and pair correlation factors. The significance of these results is discussed and the conclusions are stated in section 4.

**2.     Molecular theory and statistical mechanics approximations.**
We consider an ensemble of $N_D$ symmetric dimers consisting of two identical rigid rod-like mesogenic cores linked longitudinally at the ends of an alkyl spacer, as shown in figure 1. Such dimers can be viewed as being formed by the bonding of two 'monomers', each monomer consisting of a mesogenic core and an end-chain of half the spacer length. The mesogenic cores of the system (and the respective 'monomers') are labelled by the indices $i, j.. = 1, 2 ... N$. Since all the monomers are paired into dimers $N = 2N_D$, but the formulation that follows could be readily extended to mixtures of monomers and dimers, in which case $N > 2N_D$.

The interactions among the monomers of the system are assumed to be pair-wise additive, with the interaction energy for a pair of monomers $i, j$ denoted by $u(i, j)$. The form of $u(i, j)$ is different for bonded pairs of monomers (monomers belonging to the same dimer) and non-bonded pairs (monomers belonging to different dimers). For simplicity, *inter*-dimer core-spacer and spacer-spacer interactions are completely ignored; in other words, the spacers are assumed to merely provide the conformational constraints on the relative positions and orientations of the mesogenic cores that are attached to their ends. In this case, the interaction $u(i, j)$ for a pair of non-bonded monomers reduces to $u(i, j) = u_n(i, j)$, where $u_n$ represents the interaction energy between two non-bonded mesogenic cores and depends on the relative orientation $\omega_{ij}$ of the molecular frames attached to each of these cores and on the relative position $\mathbf{r}_{ij}$ of the origins of these frames, i.e. $u_n(i, j) = u_n(\mathbf{r}_{ij}, \omega_{ij})$. For a pair of bonded monomers we set $u(i, j) = u_b(i, j)$ where $u_b$ consists of the non-bonded core-core interaction and an additional term $u_{sp}$ associated with the conformational energy of the spacer. The relative positions and orientations of the pair of spacer-bonded mesogenic cores $i, j$ as well as the conformational energy of the spacer are fully determined given the conformational state of the spacer. The latter is specified by a set of conformational variables $v_{ij}$. Accordingly, the bonded pair potential $u_b$ is a function of the conformational variables of the spacer:

$$u_b(i, j) = u_b\{v_{ij}\} = u_n(\mathbf{r}_{i,j}\{v_{ij}\}, \omega_{i,j}\{v_{ij}\}) + u_{sp}\{v_{ij}\} \qquad . \tag{1}$$

The pair potential $u(i, j)$ enters the expression for the configurational partition function $Q$ of the ensemble of dimers in volume $V$ at temperature $T$ through the weight factor
$$G(i, j) \equiv e^{-u(i,j)/kT} , \quad (2)$$
which, depending on whether $u(i, j) = u_b(i, j)$ or $u(i, j) = u_n(i, j)$ will be denoted by $G_b$ or $G_n$ respectively, for a bonded or non-bonded pair of mesogenic cores.

The exact expression for the partition function is then
$$Q = \int d\{i\} \prod_{i \neq j} G(i, j), \quad (3)$$
with $\{i\}$ denoting collectively the set of variables describing the configurational degrees of freedom of the system. Such a set of variables can be chosen to consist, for example, of the position and orientation of one of the two cores of each dimer relative to the macroscopic frame $X, Y, Z$ and of the conformational variables $v_{ij}$ of the spacer connecting the two cores of each dimer.

Equations (2), (3) together with the corresponding expression for the configurational part of the free energy of the system
$$F = -kT \ln Q , \quad (4)$$
and the general expression for the ensemble average $<A>$ of any physical quantity $A\{i\}$ of the system,
$$<A> = (1/Q) \int d\{i\} A\{i\} \prod_{i \neq j} G(i, j) , \quad (5)$$
define the exact equilibrium statistical mechanics of the system. Clearly, however, these equations are only of formal significance. Approximations are necessary in order to actually evaluate the free energy and the relevant ensemble averages starting from the molecular interactions. When introducing such approximations it is important not to neglect or oversimplify the orientational correlations among pairs of mesogenic cores since these correlations are a crucial differentiating factor between dimer and monomer behaviour.

An approximation scheme that meets these requirements is based on the variational pair-wise decoupling of the partition function (also known as the variational two-particle cluster approximation or as the Bethe approximation [19]. Briefly, this entails introducing a variational field $\phi(i)$ and defining $\tilde{G}(i, j) \equiv G(i, j) \exp[\phi(i) + \phi(j)]$, in terms of which the exact partition function of equation (3) can be rewritten in the equivalent form,
$$Q = q^N < \prod_{i \neq j} \tilde{G}(i, j) >_0 . \quad (6)$$
Here $q \equiv \int d(i) e^{-(N-1)\phi(i)}$, and the angular brackets with the subscript 0 denote averaging with respect to the variational field distribution function $\prod_{i=1}^{N} \rho(i)$, where the single-core probability distribution $\rho(i)$ is related to the variational field according to
$$\rho(i) = e^{-(N-1)\phi(i)} / q . \quad (7)$$

The approximation consists in replacing the average in the rhs of equation (6) by a product of pair averages, namely

$$< \prod_{i \neq j} \tilde{G}(i,j) >_0 \approx \prod_{i \neq j} < \tilde{G}(i,j) >_0 \qquad . \qquad (8)$$

In this approximation the free energy of equation (4), taking into account that there are $N_D = N/2$ bonded pairs of mesogenic cores and $N(N_D - 1)$ non-bonded pairs, reduces to the expression

$$F \approx -kTN\,[\ln q + \left(\frac{1}{2}\right)\ln <\tilde{G}_b>_0 + \left(\frac{N}{2}-1\right)\ln <\tilde{G}_n>_0] \qquad . \qquad (9)$$

The approximation is completed by determining the variational field $\phi(i)$ through the functional minimisation of the above approximate expression of the free energy. The ensemble averages of quantities pertaining to a pair of monomers are evaluated by applying the pair-wise decoupling approximation of equation (8) to the expression in the rhs of equation (5). Thus, for any quantity $A_b(i,j)$ pertaining to a bonded pair of monomers we have

$$<A_b> \approx <A_b G_b>_0 / <G_b>_0 \qquad (10)$$

and a similar expression for any non-bonded pair quantity $A_n(i,j)$, by replacing $G_b$ with $G_n$.

To proceed from the above general expressions of the decoupling approximation to expression referring to the nematic phase, it is necessary first to specify the relevant variables of the variational field according to the symmetries of the phase. For rigid cores of arbitrary shape the variational probability distribution $\rho(i)$ for a single core depends only on the orientation $\omega_i$ of the core relative to the macroscopic phase-fixed frame. Denoting the orientational distribution by $f(\omega_i)$ we have, as a result of the positional uniformity of the nematic phase,

$$\rho(i) = f(\omega_i)/V \qquad , \qquad (11)$$

where $V$ denotes the sample volume. Accordingly, the variational field $\varphi(i)$ depends only on orientational variable $\omega_i$ of the core, i.e., $\varphi(i) = \varphi(\omega_i)$.

Next, certain general considerations on the form of $G_n$, $G_b$ must be taken into account. Thus, assuming that the range of the non-bonded core potential $u_n$ is of the order of the molecular dimensions we may write for $G_n$

$$G_n(\mathbf{r}_{ij}, \omega_{ij}) = 1 - \Delta(\mathbf{r}_{ij}, \omega_{ij}) \qquad , \qquad (12)$$

where the quantity $\Delta(\mathbf{r}_{ij}, \omega_{ij})$ is negligible at intermolecular distances $|\mathbf{r}_{ij}|$ exceeding the range of the interaction and therefore has significant values only within a region of $|\mathbf{r}_{ij}|$ that is of the order of the molecular dimensions. Then, according to equation (9), the relevant contribution of $G_n$ to the free energy of nematic phase comes from the orientation dependent quantity

$$\overline{\Delta}(\omega_{ij}) = (1/a^3)\int \Delta(\mathbf{r}_{ij},\omega_{ij})d\mathbf{r}_{ij} \quad , \tag{13}$$

where $a$ denotes the effective average range of the non-bonded monomer-monomer pair potential and is introduced in equation (13) in order to render the orientational function $\overline{\Delta}$ dimensionless.

Turning now to the bonded (*intra*-dimer) terms $G_b$ we note that, contrary to $G_n$, they vanish when the core-core separation exceeds the length of the fully extended spacer. In the nematic phase, only the conformationally averaged dependence of $G_n$ on the relative orientations of the mesogenic cores is of relevance. This will be denoted by $\overline{G}_b(\omega_{ij})$ and is obtained by the following formal transformation of the conformational distribution into an orientation distribution:

$$\overline{G}_b(\omega_{ij}) = \int dv_{ij} G_b(v_{ij})\delta(\omega_{ij} - \omega(v_{ij})) \quad , \tag{14}$$

where $\omega(v_{ij})$ stands for the relative orientation of the mesogenic cores when the dimer is in the conformation $v_{ij}$ and $\delta$ is the Dirac function.

On carrying out the functional minimisation of the free energy with respect to the variational field $\varphi(\omega_i)$ we obtain the following expressions for the single core distribution and for the free energy in the nematic phase:

$$f(\omega) = e^{-\xi(\omega)}/\zeta \quad , \tag{15}$$

and

$$F/NkT \approx -\ln\zeta - (1/2)\ln<\overline{G}_b>_0 + (\eta/2)<\overline{\Delta}>_0 \quad , \tag{16}$$

where

$$\zeta \equiv \int d\omega e^{-\xi(\omega)} \quad , \tag{17}$$

$$<\overline{G}_b>_0 = \int d\omega_i f(\omega_i)\int d\omega_j f(\omega_j)\overline{G}_b(\omega_{ij}) \quad , \tag{18}$$

$$<\overline{\Delta}>_0 = \int d\omega_i f(\omega_i)\int d\omega_j f(\omega_j)\overline{\Delta}(\omega_{ij}) \quad , \tag{19}$$

and $\eta = Na^3/V$ denotes the effective packing fraction of the monomers. The 'potential of mean torque', $\xi(\omega)$ in equation (15) is defined through the self consistency condition:

$$\xi(\omega_i) = \eta[\int d\omega_j f(\omega_j)\overline{\Delta}(\omega_{ij}) - <\overline{\Delta}>_0] + 1 - \int d\omega_j f(\omega_j)\overline{G}_b(\omega_{ij})/<\overline{G}_b>_0 \quad . \tag{20}$$

The inclusion of the constant terms in the rhs of this equation is merely a matter of convenience since it leads to $<\xi>_0 \equiv \int d\omega f(\omega)\xi(\omega) = 0$ and moreover it makes $\xi(\omega) = 0$ in the isotropic phase.

Calculations based on this molecular theory start with the bonded and non bonded interaction potentials $u_b$, and $u_n$ from which the orientational coupling functions $\overline{\Delta}$ and $\overline{G}_b$ are evaluated according to equations (13) and (14) and are then used to determine self-consistently the potential of mean torque from equations (15) and (20). Once $\xi(\omega)$ is determined, the free energy of equation (16) can be obtained. The ensemble average of any single-monomer quantity can then be evaluated using the orientational distribution $f(\omega)$ of equation (15). Regarding pair ensemble averages, of particular interest are the orientational averages $<\cos\theta_{ij}>$ and $<\cos\theta_i \cos\theta_j>$, with $\theta_{ij}$ denoting the angle formed by the long axes of the mesogenic cores of monomers *i* and *j* and $\theta_i, \theta_j$ denoting the angle of each such axis with the director of the phase. For a bonded pair of monomers $<\cos\theta_{ij}>$ is given, according to equation (10), by the expression

$$<\cos(\theta_{ij})_b> \approx \int d\omega_i d\omega_j f(\omega_i) f(\omega_j) \cos(\vartheta_{ij})_b \overline{G}_b(\omega_{ij}) / <\overline{G}_b>_0 \quad . \tag{21}$$

For a non-bonded pair, the respective expression is

$$<\cos(\theta_{ij})_n> \approx -(\eta/N) \int d\omega_i d\omega_j f(\omega_i) f(\omega_j) \cos(\vartheta_{ij})_n \overline{\Delta}(\omega_{ij}) \quad , \tag{22}$$

where account has been taken of $<G_n>_0 = 1$ and of the a-polarity of the nematic phase, as a result of which $\int d\omega_j f(\omega_j) \cos(\theta_{ij})_n = 0$. Analogous equations hold for the ensemble averages $<\cos\theta_i \cos\theta_j>$ of bonded and non-bonded pairs.

To focus on how the spacer influences the nematic behaviour of the dimer relative to that of the monomer in the absence of other structural complexities or asymmetries, we shall restrict our attention to uniaxial nematic phases and consider alkyl spacers connecting perfectly uniaxial mesogenic cores bearing permanent dipole moments that are directed along the symmetry axis ('long' axis) of the cores. We shall also assume that the core symmetry axis is collinear with the bond at the core-spacer linkage. In this case, the orientational couplings $\overline{\Delta}$ and $\overline{G}_b$ for a pair of cores *i, j* are functions of the angle $\theta_{ij}$ between the core long axes while the potential of mean torque is a function of the angle $\theta$ between the symmetry axis of the mesogenic core and the nematic director. It is then straightforward to formulate the free energy in terms of order parameters by expanding the orientational couplings $\overline{\Delta}(\theta_{ij})$, $\overline{G}_b(\theta_{ij})$ and the potential of mean torque $\xi(\theta)$ in series of Legendre polynomials $P_l(\cos\theta)$. The ensemble averages of these polynomials define the mesogenic core orientational order parameters of tensor rank *l*, namely

$$<P_l>_0 \equiv \int f(\theta) P_l(\cos\theta) d\cos\theta \quad . \tag{23}$$

The details of this expansion are given in the appendix.

In general, the Legendre series representation introduces an infinite number of expansion coefficients as well as order parameters of unlimitedly high ranks. However, the usefulness of such representation in actual calculations lies in the possibility of obtaining fairly accurate results by truncating the expansion to some computationally reasonable rank. The truncation is justified by the rapidly decreasing magnitude of the orientational order parameters with increasing rank in all common, low molar mass, nematic fluids and by the possibility of adequately representing the dominant intermolecular interactions by means of a fairly small number of low rank expansions terms. Even so, the actual calculations are not particularly facilitated by using the Legendre series formulation and in fact, the results presented in the next section are obtained by solving the self consistency conditions of equations (20) for the full orientational couplings rather than for their low rank representatives in the Legendre series expansion. These results are then used to assess the accuracy of the results obtained with the truncated series.

Finally, it should be noted that the derivation of the free energy within the variational pair-wise decoupling approximation includes the description of the monomer system as a limiting case. Indeed, on removing the connected pair interactions (by setting $G_b = 0$) from equations (16), (20) one obtains the free energy and potential of mean torque for the monomer system at effective packing fraction $\eta$.

### 3.    Calculations with model interactions.

As is well known, the use simple interaction potentials is rarely adequate for a quantitative description of mesophases on the molecular scale and moreover it is in no way established that a quantitative description can be achieved by means of exclusively pair-wise additive interactions [20, 21] irrespectively of their degree of complexity. With that in mind, and given the pair-wise decoupling involved in the free energy and the ensemble averages, the scope of the present calculation is simply to rationalize the nematic behaviour of the dimers in relation to that of the monomer systems within the context of a simple molecular theory that could be successful in predicting qualitative trends.

In that context we have used the potential previously employed in the molecular simulations of dimers by M. Wilson [16]. Thus, the dimer is treated as a collection of anisotropic dipolar Gay-Berne segments, corresponding to the mesogenic cores, and spherical Lennard-Jones sites, corresponding to the methylene groups of the alkyl-spacer. The non-bonded potential is therefore written as

$$u_n(i,j) = U_{ij}^{GB} + U_{ij}^{Dipolar} \qquad , \tag{24}$$

with $U_{ij}^{GB}$ denoting the Gay-Berne potential, parameterized as in ref[ ], and the dipole-dipole interaction given by

$$U_{ij}^{Dipolar} = \mu^2 \frac{\cos\theta_{ij} - 3(\mathbf{r}_{ij} \cdot \mathbf{\mu}_i)(\mathbf{r}_{ij} \cdot \mathbf{\mu}_j)}{r_{ij}^3} \qquad , \tag{25}$$

where $\mathbf{\mu}_i$ the direction of the mesogenic dipole of strength $\mu$ and $\mathbf{r}_{ij}$ is the direction of vector that connects the dipoles $i, j$. The numerical value $\mu = 2$ Debye has been used for the dipole moment magnitude in the present calculations.

The corresponding orientational coupling $\overline{\Delta}(\theta_{ij})$ is plotted in figure 2a, from where it is apparent that its angular dependence is essentially conveyed by a $P_2(\cos\theta_{ij})$ behaviour, with some small asymmetry ($P_1(\cos\theta_{ij})$ contribution) due to the attachment of permanent longitudinal electric dipole moments on the mesogenic cores.

To model the flexible spacer we have assumed constant bond lengths, $L_{C-C} = 1.53\,\text{Å}$ and bond angles, $\theta_{CCC} = 112°$. With these constraints, the conformations of the dimmer are specified completely by the set of the dihedral angles, $\{\varphi_i\}$ and we have for the bonded-pair potential

$$u_b(i,j) \equiv u_b\{\varphi\} = U_{ij}^{GB} + U_{ij}^{Dipolar} + U^{Spacer}\{\varphi\} \quad , \tag{26}$$

where for $U_{ij}^{GB}$ and $U_{ij}^{Dipolar}$ we use the same forms as for the non-boded interactions and the conformational energy of the spacer, $U^{Spacer}\{\varphi\} = U^{diherdal}\{\varphi\} + U^{LJ}\{\varphi\}$ includes the torsional potential of the C-C bonds, $U^{diherdal}\{\varphi\}$, modeled as a Ryckaert-Bellemans [17] expansion and $U^{LJ}\{\varphi\}$ corresponds to site-site Lenard-Jones interactions between non bonded sites of the spacer with the same parameterization as in [16].

The respective orientational coupling $\bar{G}_b(\theta_{ij})$ is plotted in figure 2b, for an odd- and an even-carbon number spacer at the same temperature. Noteworthy is the marked qualitative difference in the angular dependence for these two spacers and also that, in both cases this dependence is essentially conveyed by a $P_3(\cos\theta_{ij})$, albeit with opposite sign for odd and even spacers. The same function, $\bar{G}_b(\theta_{ij})$, was evaluated using the 3-state RIS scheme [9], instead of the semiatomistic model described above, to generate the spacer conformations. The results of these calculations are shown the bar graph of figure 2c. For the RIS parametrization we have used constant bond lengths, $L_{C-C} = 1.53$ Å and bond angles, $\theta_{CCC} = 112^\circ$ and energy parameters $E_{tg}/kT = 1$, $E_{g^\pm g^\mp} = 6E_{tg}$, torsion angle minima at $112.5^\circ$ and rejection of all self-intersecting conformations. For the detection of the latter conformations, the methyl groups of the spacer are represented by hard spheres of diameter 3.5Å and the mesogenic cores by hard cylinders with length 18Å and width 6.6Å, in accord with the dimensions of a cyano-biphenyl core. These calculations were repeated, this time allowing for a uniform distribution of the torsion angle within a range of $\pm 5^\circ$ centered at the RIS minima [18]. The results are shown on the solid-line plot of fig 2c. As it is clear the dominance of the $P_3(\cos\theta_{ij})$ on the behaviour of the orientational coupling $\bar{G}_b(\theta_{ij})$ persist in all three different model calculations indicating that this feature of the orientational coupling of the bonded mesogens is an intrinsic property of that type of dimmers and not the circumstantial outcome of a particular modeling.

Using the orientational coupling functions obtained by the semi-atomistic modelling of the dimers we have calculated the nematic-isotropic transition temperatures, expressed in units of the Gay-Berne interaction parameter $\varepsilon_0^{GB}/k$ [16], as a function of spacer length (Figure 3a), and the respective values of the core order parameter $<P_2>_0$, at the transition (Figure 3b). For the above calculations we have assumed constant packing fraction of the mesogenic cores for the monomer system and for all the dimers irrespectively of their specer length. This is a crude approximation and the results should be viewed accordingly, particularly in view of the sensitivity of the phase transition calculations to variations of the packing fraction $\eta$. It is apparent from Figures 3a,b that the strong odd-even effect for the short dimers disappears for the longer members since, the high number of accessible conformations for dimers with long spacers, weaken significantly the orientational coupling between the bonded mesogen pairs.

The dipolar correlation factors are quantities of particular significance to the static and low frequency dielectric properties of both the monomer and the dimeric systems. In particular, the ensemble average that determines the orientational polarisation part of the dielectric permittivity is the total dipole tensor $<M_A M_B>$, where $A,B = X,Y,Z$ are component indices referring to a macroscopic axis frame $X,Y,Z$ and $M_A$ denotes the $A$ component of the total dipole moment of the system, $M_A = \sum_{i=1}^{N} \mu_A(i)$.

Taking into account that the dipolar moments are paired into N/2 dimers we have

$$<M_A M_B> = N<\mu_A \mu_B> + N<\mu_A \mu_B'>_{bonded} + N(N-2)<\mu_A \mu_B'>_{non-bonded}$$

Moreover, in a uniaxial phase, with $Z$ chosen to be the symmetry axis (director) of the phase, only the diagonal components $(A=B)$ of the dipole tensor are non-vanishing. In the present case of purely longitudinal dipoles, these components can be evaluated according to the following relations

$$<\mu_Z \mu_Z> = \mu^2 <\cos^2\theta>$$
$$<\mu_X \mu_X> = <\mu_Y \mu_Y> = \mu^2(1-<\cos^2\theta>)/2$$
$$<\mu_Z \mu_Z'>_{bonded} = \mu^2 <\cos\theta \cos\theta'>_b$$
$$<\mu_X \mu_X'>_{bonded} = <\mu_Y \mu_Y'>_{bonded} = \mu^2(<\cos\theta_{\mu\mu'}>_b - <\cos\theta \cos\theta'>_b)/2$$

with nalogous expressions holding for the non-bonded dipole pairs and with $\theta_{\mu\mu'}$ denoting the angle between the directions of dipoles $\mu$ and $\mu'$. Thus in the present, fully uniaxial case, all the dipolar correlation factors are conveyed by the pair averages $<\cos\theta_{ij}>$ and $\langle\cos\theta_i \cos\theta_j\rangle$. These can be evaluated according to equations (21) and (22). For the monomeric system there is just one pair average, pertaining to non-bonded mesogenic cores. The temperature dependence of this average is shown in figure 4 together with the orientational order parameter $\langle P_2 \rangle$. It is apparent from the plots in fig 4 for the monomeric system that there is a tendency for antiparallel dipolar association ($\langle\cos\theta_{ij}\rangle < 0, \langle\cos\theta_i \cos\theta_j\rangle < 0$) of the molecular dipoles in both, the nematic and the isotropic phase. However, in the isotropic phase the dipolar associations are rather weak and almost insensitive in temperature changes. At the NI transition temperature the dipolar factors jump to much higher values and increase almost linearly on lowering the temperature.

For dimers there are two additional pair averages, associated with the bonded pairs. The temperature dependence of these averages, together with the respective averages of non-bonded pairs are shown on fig 4a-d for four different dimers with spacer lengths ranging from seven up to ten carbon atoms. It is apparent that the temperature dependence of the orientational correlation factors of bonded mesogenic pairs differ appreciably for spacers of opposite parity while the correlation factors of non-bonded pairs exhibit the same temperature dependence and barely differ from their counterparts of the monomeric system. The negative values for the correlation factors of bonded mesogen pairs s in the nematic phase of even dimers reflect a favouring of linear dimer conformations that bring the mesogenic dipoles antiparallel. This tendency grows stronger on lowering the temperature. The situation changes for odd dimers where the corresponding averages near the transition temperature are much lower but tending to positive values on lowering the temperature. This reflects the favouring of parallel dipolar association, indicating that the dimer tends towards to U-shaped conformations. The interplay between bonded and non-bonded correlation factors as well as the qualitative dependence of that interplay on the spacer parity can account for the diverse trends observed in experimental dielectric studies for the permittivity of nematic dimers [22-25] and some of the marked differences from the behaviour of the monomer compounds. Thus for example, the experimentally observed decrease of the parallel permittivity component $\varepsilon_{//}$ with decreasing temperature (increasing orientational order) and, more notably, its initial increase and subsequent decrease are readily accounted for by the competing tendencies in the temperature dependence of the order parameters and correlation factors of bonded and non bonded pairs shown in figures 4 and 5.

## 4. Discussion and conclusions

This is, to our knowledge, the first time that a molecular theory of liquid crystal dimers starting from explicit inter- and intra-molecular interactions is presented. The application of the theory to the description of the nematic phase has successfully accounted for the odd-even alternations at the phase transition and for the parity and temperature dependence of dipolar correlation factors of dimers, while singling out the differentiating factors from the monomer behaviour.

The statistical mechanics formulation of the theory is not restricted in any way to the nematic phase and this makes it possible to use it for the description of smectic polymorphism in LC dimers. Previously existing molecular theories [10-12], have also been successful in accounting for some of the basic features of the nematic behaviour of dimers. However, these theories, being based on some form of effective interaction of the dimer molecule with its nematic environment, introduced through a phenomenological potential of mean torque, can simply not be extended to smectic ordering. Further, these phenomenological treatments are intrinsically limited to tensor interactions of second rank, a limitation that in view of the present results could be more severe than initially expected, and moreover the mean-field nature of these phenomenological theories can only give trivial results for molecular pair correlations. In contrast, the results presented here show that by properly accounting for the orientational correlations between bonded and non-bonded pairs of mesogenic units, a proper theoretical framework for rationalising the dielectric properties of nematic dimers is obtained, while maintaining all the successful features of phenomenological mean-field treatments.

**Acknowledgments**
PKK acknowldges support from the Ministry of Education-Hellas through a research fellowship in the framework of the project "Pythagoras" and AGV acknowledges support through the "Caratheodores" research programme of the University of Patras.

**Appendix**

Here we provide an equivalent representation of the free energy and the potential of mean torque of eqs(16) and (20) respectively in terms of Legendre polynomials $P_l(\cos\theta)$ and the corresponding orientational order parameters $<P_l>_0$ defined in equation (23). The representation is based on the Legendre series expansion of the orientational couplings,

$$\overline{G}_b(\theta_{ij}) = \sum_{l=0}^{\infty} \overline{G}_b^{(l)} P_l(\cos\theta_{ij}) \quad , \tag{A1}$$

$$\overline{\Delta}(\theta_{ij}) = \sum_{l=0}^{\infty} \overline{\Delta}^{(l)} P_l(\cos\theta_{ij}) \quad . \tag{A2}$$

The expansion coefficients $\overline{G}_b^{(l)}$, $\overline{\Delta}^{(l)}$, together with the coefficients $\xi^{(l)}$ of the respective expansion of the potential of mean torque,

$$\xi(\theta) = \sum_{l=1}^{\infty} \xi^{(2l)} P_{2l}(\cos\theta) \quad , \tag{A3}$$

can be used to write we the nematic-isotropic free energy difference in terms of the order parameters as follows,

$$(F_{iso} - F_{nem})/NkT = \ln(\frac{1}{2}\int \exp[-\sum_{l=1}^{\infty} \xi^{(2l)} P_{2l}(\cos\theta) d\cos\theta]) + \\ + \frac{1}{2}\ln(1 + \sum_{l=1}^{\infty} \frac{\overline{G}_b^{(2l)}}{\overline{G}_b^{(0)}} <P_{2l}>_0^2) - \frac{\eta}{2}\sum_{l=1}^{\infty} \overline{\Delta}^{(2l)} <P_{2l}>_0^2 \tag{A4}$$

According to the self consistency equation (20) the potential of mean torque $\xi(\theta)$ is related to the coefficients $\overline{G}_b^{(l)}, \overline{\Delta}^{(l)}$ and the order parameters $<P_l>_0$ as follows,

$$\xi(\theta) = \sum_{l=1}^{\infty} (P_{2l}(\cos\theta) - <P_{2l}>_0) <P_{2l}>_0 \left[ \eta\overline{\Delta}^{(2l)} - \frac{\overline{G}_b^{(2l)}}{\sum_{l'=0}^{\infty} \overline{G}_b^{(2l')} <P_{2l'}>_0^2} \right]$$

Due to the assumed apolarity of the nematic phase, only even rank expansion coefficients and order parameters appear in the *l*-summations of the expressions for the free energy and the potential of mean torque. It should be noted however, that the expansions of the orientational couplings in equations (A1-2) do include odd-ranked terms and in fact, the odd-rank contributions to the bonded coupling, particularly the *l=3* terms, seem to be quite significant according to the results in figure 2b.

It is of some interest to consider the form of the free energy and of the potential of mean torque in the limit of very weak spacer-mediated orientational correlations of the mesogenic cores, as it is the case with very flexible or very long spacers. In this limit $\left|\frac{\overline{G}_b^{(2l)}}{\overline{G}_b^{(0)}}\right| << 1$ and the expression for the free energy difference of eq (A4) reduces to

$$(F_{iso} - F_{nem})/NkT = \ln(\frac{1}{2}\int \exp[-\sum_{l=1}^{\infty}\xi^{(2l)}P_{2l}(\cos\theta)d\cos\theta]) + \frac{1}{2}\sum_{l=1}^{\infty}\lambda^{(2l)}<P_{2l}>_0^2 \quad , \quad (A5)$$

where

$$\lambda^{(l)} = \frac{\overline{G}_b^{(2l)}}{\overline{G}_b^{(0)}} - \eta\overline{\Delta}^{(2l)} \quad . \tag{A6}$$

The even-rank expansion coefficients of the potential of mean torque of eq (A5) are in this case given by

$$\xi^{(2l)} = -\lambda^{(2l)}<P_{2l}>_0 \quad . \tag{A7}$$

It is apparent from equations (A5) and (A7) that, in the limit of a "loose" spacer, the dimers are formally equivalent to a system of monomers interacting via an effective pair potential whose Legendre series expansion is determined by the coefficients $\lambda^{(l)}$ of equation (A7) which include both, the direct core-core interactions ($\overline{\Delta}^{(l)}$ terms) and the spacer mediated interactions ($\overline{G}_b^{(l)}$ terms).

On further assuming that the effective interaction is such that the terms $\lambda^{(l)}$ of rank $l > 2$ can be ignored, the resulting potential of mean torque and free energy difference become formally identical to those of the Maier-Saupe (MS) theory of nematics [26]. In this sense, the free energy formulation of equation (A4), when restricted to terms of rank $l \le 2$ is an extension of the MS theory to nematic dimers. However, there are two important points of difference between this restricted form of the present theory of dimers and the MS theory:

(a) the coefficients $\lambda^{(l)}$ are explicitly related to the core-core potential, both direct and through-spacer, according to equation (A6) and

(b) the mean-field character of the MS theory cannot address orientational correlations between pairs of mesogenic cores.

In addition, it should be kept in mind that, according to the results in figure (2), while the neglect of the tensor coefficients $\overline{\Delta}^{(l)}$ with $l > 2$ could be acceptable, since normally these are considerably smaller than the dominant $\overline{\Delta}^{(2)}$ term, the same is not true for the $\overline{G}_b^{(l)}$ coefficients where rank $l = 3$ coefficients could be more significant than those of rank $l = 1,2$, even for fairly long spacers. Thus a MS-type of treatment of nematic dimers could not be expected to adequately account for some of the essential features of their nematic ordering.

# Figure Captions

**Figure 1.** Schematic representation of a symmetric even dimmer with six metlhylene sites. The mesogenic groups are assumed perfectly uniaxial carrying a central longitudinal point dipole **μ**.

**Figure 2.** Plots of the orientation dependence of the interactions.
- (a) orientational coupling $\bar{\Delta}(\theta_{ij})$ for non-bonded mesogenic cores.
- (b) normalized orientational coupling $\bar{G}_b(\theta_{ij})$ between mesogenic units belonging the same dimer. Results are shown for two spacers of lengths 11 (solid line) and 12 (dashed line) methylene groups.
- (c) same as (b) for a dimer with spacer of 11 methylens. The bar-diagram results are obtained for the alkyl spacer modeled according to the 3-state RIS approximation. The continuous line corresponds to the same RIS parametrization but allowing uniform distribution of the torsion angles within a range of ±5° centered at the RIS minima.

**Figure 3.** Plots of the spacer-length dependence of the transition transition temperatures (a) and orientational order parameter at the transition (b) for symmetric dimers. The continuous lines on the plots indicate the transition temperature of the monomeric system and the respective value of the order parameter at the transition. All the results are obtained at the same value of the packing fraction of the mesogenic units, $\eta=0.7$, for the monomer and the dimers irrespectively of their spacer length.

**Figure 4.** Temperature dependence of dipole correlation factors $\langle \cos(\theta_{ij})_n \rangle$, $\langle \cos(\theta_i)_n \cos(\theta_j)_n \rangle$ and orientational order parameter $\langle P_2 \rangle$ for the monomer system.

**Figure 5.** Temperature and spacer length dependence of dipole correlation factors $\langle \cos(\theta_{ij})_{n(b)} \rangle$, $\langle \cos(\theta_i)_{n(b)} \cos(\theta_j)_{n(b)} \rangle$ for bonded and non-bonded mesogenic pairs.

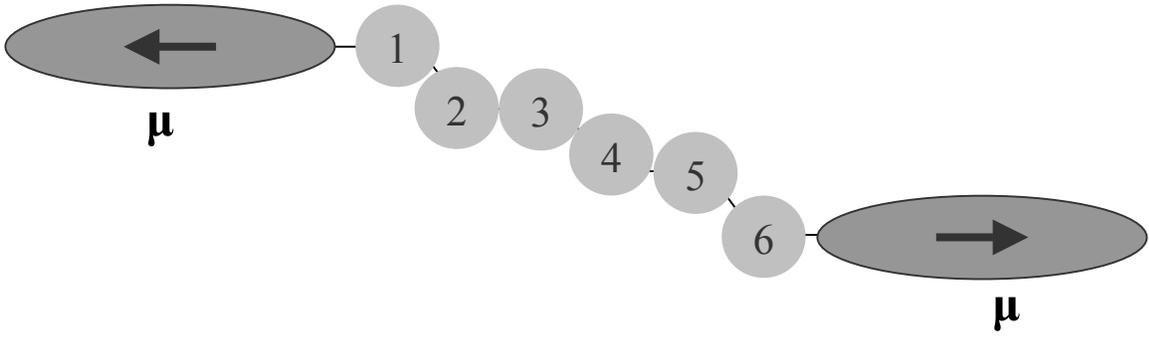

**FIGURE 1**

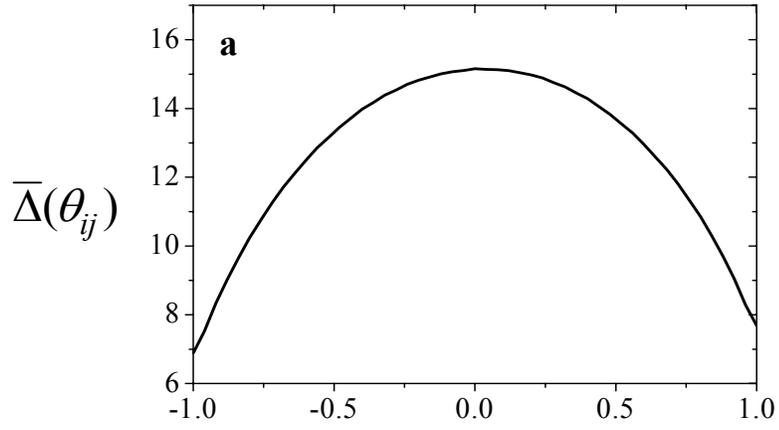
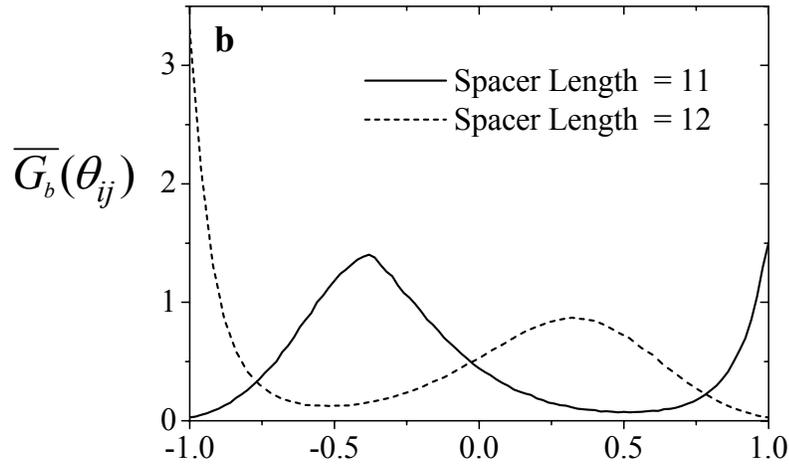
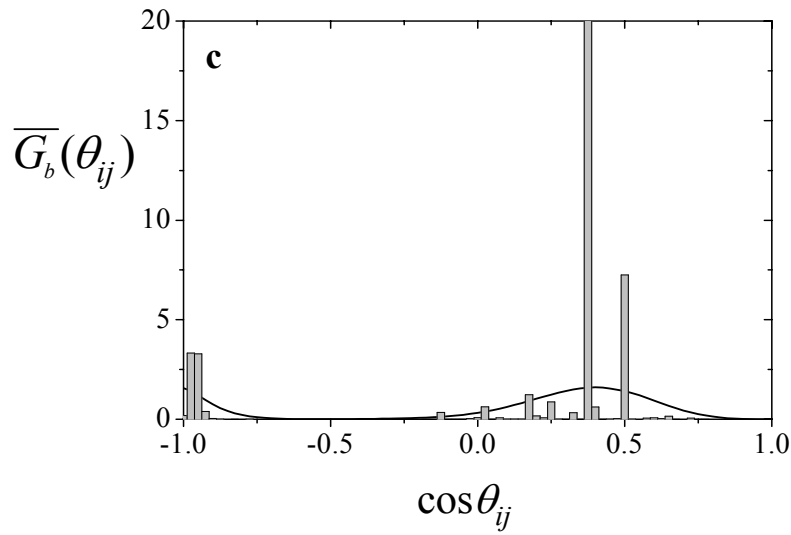

**FIGURE 2**

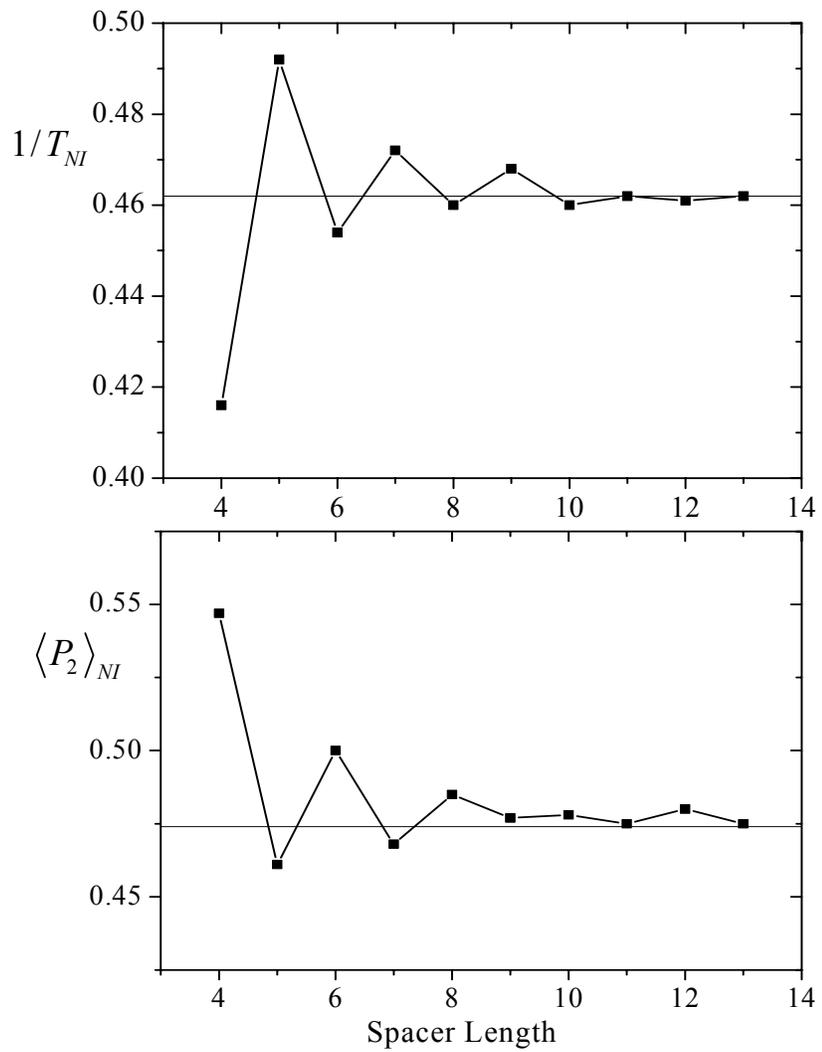

**FIGURE 3**

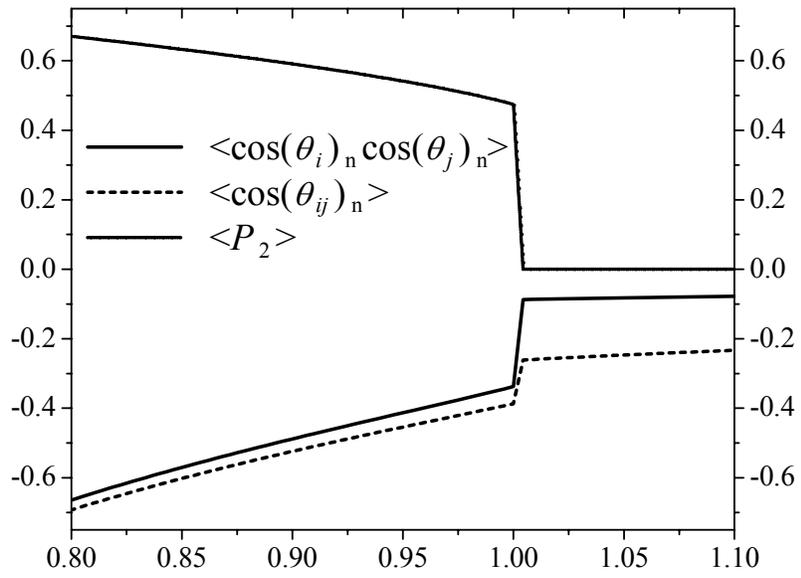

**FIGURE 4**

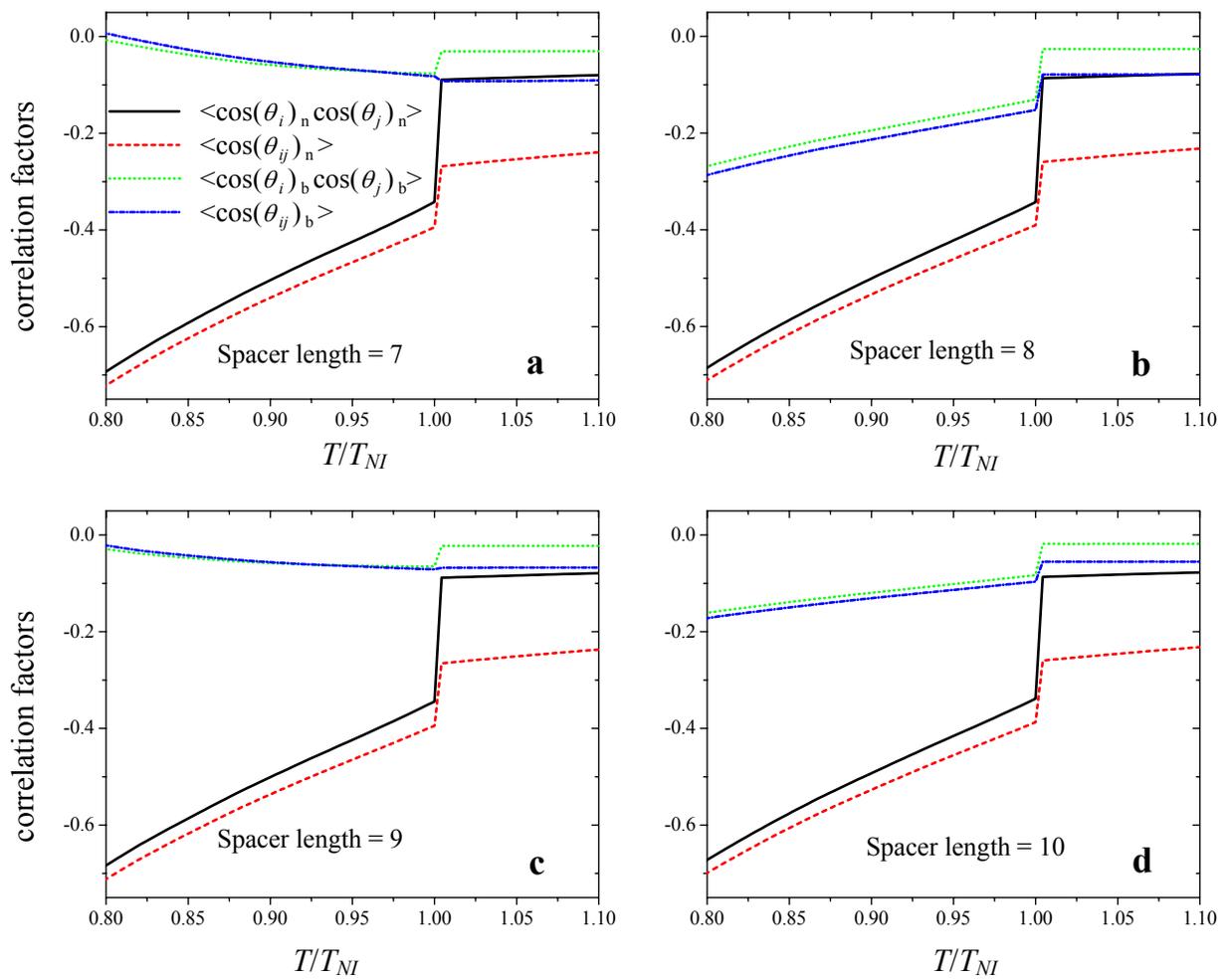

**FIGURE 5**